\documentstyle[12pt]{article}

\def\Bbb{\bf}

\def\dir{\hat\nabla}
\def\tm{{\tilde m}\vphantom{m}}
\def\tn{{\tilde n}\vphantom{n}}
\newcommand{\stack}[2]
 {\stackrel{\scriptstyle #1}{#2}\hspace{-3.5pt}\vphantom{#2}}

\title{Tensionless  strings \\ and the space-time signature change}
\author{S.A. Pol'shin$^{1}$\thanks{E-mail: itl593@online.kharkov.ua}
\  and \ A.A. Zheltukhin$^{2}$\thanks{E-mail: aaz@physto.se} \\
\normalsize $^{1}$Kharkov State University \\
\normalsize 310077, Svobody Sq. 4, Kharkov, Ukraine \\ \\
\normalsize $^{2}$NSC Kharkov institute of Physics and Technology \\
\normalsize 310108 Kharkov, Ukraine \\
\normalsize and Institute of  Theoretical Physics , University of Stockholm\\
\normalsize Box 6730, S-11385 Stockholm, Sweden}
\date{}
\begin{document}
\maketitle
\begin{abstract}
We construct a tensionless string model in a four-dimensional space-time
 ${\bf R}^{2,2}$ with the (2,2) signature and solve the string equations.
 We find that the signature change radically changes the structure of the
 holonomy group of the null worldsheet. As a result the introduction of
the worldsheet Dirac operator invariant under the holonomy  group becomes
possible. We show that the such possibility is absent in the tensionless
string model in the (3+1)-dimensional Minkowski space.
 \end{abstract}

The extended tensionless objects - null strings~\cite{Shi} and null
p-branes ~\cite{Zpb} are characterized by a simple dynamics which
combines the properties of massless particles (equations of motion) and
strings/p-branes (constraints). It gives a possibility to probe the
complicated problems of the nonlinear (super)string/p-brane dynamics in the
simplified picture of the  null (super)string/p-brane
dynamics~\cite{null-7,Zpb}. In particular, null (super)string/p-brane may be
covariantly quantized~\cite{Ba-Zh} and their equations are exactly  solvable
for the cases of some interesting  backgrounds~\cite{Il-Zh}. It was shown in
\cite{Ro-Zh} that the  null string action may be used as the zero
 approximation action in a perturbative description of the string dynamics in curved
spaces~\cite{deV-Sa}. In \cite{Tw1,BLT} was observed that the inclusion of
the worldsheet electromagnetism  into the null superstring action allows a
geometric interpretation of the Green-Schwarz superstring symmetries and
leads to the string tension generation.
The examination of the zero tension limit in the D-brane theory~\cite{Un-Ul}
allows to obvserve its parton-like structure which is
close to the structures of the matrix model of M-theory~\cite{Bank-} and
the p-brane dynamics~\cite{Yang-}. It was shown in ~\cite{SGNSW,BT} that
the tension generation mechanism  \cite{Tw1} is also selfconsistent with
the D-brane dynamics.

But there is one important difference between the tensionless and tensile
 superstrings. The tensionless (super)strings/p-branes have no any
critical dimension and are selfconsistent objects, in particular, in the
(3+1)-dimensional Minkowski space. Therefore, it seems rather interesting
to investigate  the reaction of the  null string dynamics under the
deformations of the space-time structure. Topological deformations belong
to the most interesting ones, because they affect the string dynamics and
may change the value of the critical dimension. The latter possibility
was realized in the spinning string model with the $N=2$ worldsheet conformal
symmetry~\cite{Vech-} which turned out to be selfconsistentl in a
(2+2)-dimensional space-time~\cite{Fra}. The massless fields  generated by
the  $N=2$ spinning string are the self-dual Yang-Mills or gravitation
fields~\cite{null-1}, which  are  closely connected with the exactly
integrable systems~\cite{Bel-}. A supersymmetric generalization of the
self-dual  Yang-Mills/supergravity theories in (2+2) dimensions  has shown
the existence of the real chiral scalar supermultiplet~\cite{Ket1}.
Therefore, one can conclude that the signature changing is a constructive operation  which shed light on the subtle correllation between the world-sheet symmetry,
 supersymmetry, critical dimension and exact integrability.

The objective of the present letter is to continue the investigation of
 posible physical effects in the string dynamics which arise from the
signature change. To this end we construct here a model of free bosonic null
 string moving in the flat space ${\bf R}^{2,2}$ and prove the exact
 solvability of the string equaitons. We establish that the signature change
 opens a possibility to build the worldsheet Dirac operator invariant under
 the holonomy group of the null worldsheet. We show that the such possibility
  is absent in the case of null string embedded into the Minkowski space
 ${\bf R}^{3,1}$.

Our rules for the indices usage are the following:
$\mu,\nu,\ldots=1,\ldots, 4$ over ${\Bbb R}^{2,2}$ and $0,\ldots, 3$ over
${\Bbb R}^{3,1}$; $i,k=1,2$; $a,b,\ldots=1,\ldots,3$; the worldsheet
indices
$\alpha,\beta,\ldots=1,2$.

\paragraph{1}
The symmetry group of the ${\bf R}^{2,2}$ space is the semidirect product
 of the Abelian translation subgroup with the generators $P_\mu$ and the
$SO(2,2)$ group of the orthogonal transformations with the generators
 $J_{\mu\nu}=-J_{\nu\mu}$:
\begin{equation}\label{null1}
\begin{array}{l}
{[}P_{\mu}, P_{\nu}{]}= 0 \\
{[}P_{\mu}, J_{\nu \rho}{]}=\eta _{\mu \nu} P_{\rho}-
\eta _{\mu \rho}P_{\nu} \\
{[}J_{\mu \nu}, J_{\rho \sigma}{]}=\eta _{\mu \sigma}J_{\nu \rho}+
\eta _{\nu \rho}J_{\mu \sigma}-\eta _{\mu \rho}J_{\nu \sigma}-
\eta _{\nu \sigma}J_{\mu \rho},
\end{array}
\end{equation}
where $\eta_{\mu\nu}=\mbox{diag}(+1,+1,-1,-1)$ is the metric tensor.
The matrix elements of the  $SO(2,2)$ vector representation are
\begin{equation}\label{null2}
(J^{(v)}_{\mu\nu})^{\rho}_{\ \sigma}=
\delta^{\rho}_{\mu}\eta_{\nu\sigma} - \delta^{\rho}_{\ \nu}\eta_{\mu\sigma}.
\end{equation}
The null plane having the coordinates $x^\mu$ and passing through the
 coordinate system origin in ${\bf R}^{2,2}$ is given by
 $x\cdot m_{i}=0,\ i=1,2$, where
$$m_{1}^{\mu}=(0,-1,0,1) \quad,\quad m_{2}^{\mu}=(1,0,-1,0).$$
Then one can express the couple of the coordinates $\zeta^i$, parametrizing
 this null plane, by the equations $\zeta^{i}=n_{i}\cdot x$, where
\begin{equation}\label{null4}
n_{1}^{\mu}=(1, 0, 1, 0) \quad,\quad n_{2}^{\mu}=(0, 1, 0, 1)
\end{equation}
$$n_{i}\cdot n_{k}=m_{i}\cdot m_{k}=0 \quad,\quad
n_{i}\cdot m_{k}=2\varepsilon_{ik}.$$

 Let us define now the subgroup of the rotations group transforming the null
plane onto itself by the conditions
 \begin{equation}\label{null5}
L^{\mu}_{\ \nu}n^{\nu}_{i}=\alpha_{ik}n^{\mu}_{k},
\end{equation}
where $L^{\mu}_{\ \nu}$ are the  $SO(2,2)$ global transformations and
 $\alpha_{ik}$ are some coefficients depending on $L_{\mu\nu}$. One can
write down the infinitesimal transformations with the parameters
$\Delta\omega_{\mu\nu}=-\Delta\omega_{\nu\mu}$ as
\begin{equation}\label{null6}
L^{\mu}_{\ \nu}=\delta^{\mu}_{\nu}+
\frac{1}{2}\Delta\omega^{\rho\sigma}(J_{\rho\sigma}^{(v)})^{\mu}_{\ \nu}=
\delta^{\mu}_{\nu}+\Delta\omega^{\mu}_{\ \nu}.
\end{equation}
The substitution of the representation~(\ref{null6}) into Eqs.(\ref{null5})
  together with the use of the explicit form of $n_i$ restrict the possible form of $\Delta\omega_{\mu\nu}$. As a result we find $L_{\mu\nu}$  in the form of a linear combination of the unity element
and the symmetry group generators of the null plane. It is easy to check
that these generators  are presented in the form
\begin{equation}\label{null6a}
\begin{array}{l}
2X_{1}=J_{12}-J_{34} \quad, \quad 2X_{2}=J_{13}-J_{24} \quad,\quad
2X_{3}=J_{14}+J_{23} \\
D=J_{13}+J_{24} \quad,\quad S=J_{12}-J_{14}+J_{23}+J_{34}
\end{array}
\end{equation}
and  satisfy the  following commutation relations
\begin{equation}\label{null7}
[X_{a},X_{b}]=\varepsilon_{abc}X^{c} \quad,\quad
{[}X_{a},S{]}={[}X_{a},D{]}=0 \quad,\quad {[}D,S{]}=2S,
\end{equation}
where the three-dimensional indices are raised and lowered by means of the
tensor \linebreak $\rho_{ab}=\mbox{diag}(-1,+1,+1)$ and $\varepsilon_{1
23}=1$.
It is easy to  find  that Eqs.~(\ref{null7}) produce the direct product of
 the $SO(2,1)$ group and the solvable group with the generators $S$ and $D$.
Let us consider now the action of these groups onto the vectors $n_i$:

1) The generator $S$ does not change the vectors $ n_{i}$ of the null plane:
$(S^{(v)})^{\mu}_{\ \nu}n^{\nu}_{i}=0$ and may be omitted without any
 loss.

2)  $D$ is the dilatation generator: $(D^{(v)})^{\mu}_{\ \nu}n^{\nu}_{i}=
n^{\mu}_{i}$ and the Abelian group created by it will be denoted by  $\cal D$.

3) The action of the rest generators  $X_a$ is given by
 \begin{equation}\label{null8}
(X^{(v)}_{a})^{\mu}_{\ \nu}n_{i}^{\nu}=(\sigma_{a})_{ik}n_{k}^{\mu},
\end{equation}
where the matrices
$$\sigma_{1}=\left(
\begin{array}{rr}
0 & 1 \\
-1 & 0
\end{array}\right) \quad,\quad
\sigma_{2}=\left(
\begin{array}{rr}
-1 & 0 \\
0 & 1
\end{array}\right) \quad,\quad
\sigma_{3}=-\left(
\begin{array}{rr}
0 & 1 \\
1 & 0
\end{array}\right)$$
are the Pauli matrices in the 2+1 dimensional space:
\begin{equation}\label{null8a}
\sigma_{a}\sigma_{b}=\rho_{ab} -\varepsilon_{abc}\sigma^{c}.
\end{equation}
Taking into account the relations $\varepsilon^{ik}=-(\sigma_{1})_{ik}$
($\varepsilon^{12}=-\varepsilon^{21}=1$), the symmetry property of the
  matrices $\sigma_{1}\sigma_{a}$ and the transformation rules (\ref{null8})
  we find that the bilinear form $\varepsilon^{ik}{n'}_i^\mu
{n'}_k^\nu=\varepsilon^{ik}{n}_{i}^{\mu}{n}_{k}^{\nu}$ remains  invariant
 under these transformations. Then, in view of the well known isomorphism
 $SO(2,1)\simeq Sp(1,{\Bbb R})$ one can consider the null-plane as a flat symplectic manifold.
  To yield  the complete symmetry group of the null plane we should extend
  $Sp(1,{\bf R})$ by the generators
\begin{equation}\label{null9}
\Pi_{i}=P_{i}-P_{i+2},
\end{equation}
which produce the shifts of the null plane into itself.

Using~(\ref{null1})  we find that $\Pi_{i}$ compose the set of the $Sp(1,
{\bf R})$ and $\cal D$ tensor operators:
\begin{equation}\label{null10}
{[}\Pi_{i},X_{a}{]}=(\sigma_{a})_{ik}\Pi_{k} \quad,\quad
{[}\Pi_{i},D{]}=-\Pi_{i}.
\end{equation}

\paragraph{2}

The Dirac operator over ${\Bbb R}^{2,2}$ reads
\begin{equation}\label{null12}
\dir=\gamma^{\mu}P_{\mu},
\end{equation}
where $P_{\mu }=\frac{\partial}{\partial x^{\mu}}$ are the translation
generators and $\gamma$ are the Dirac matrices with  the algebra
 $\{\gamma_{\mu},\gamma_{\nu}\}=2\eta_{\mu\nu}$. The Dirac operator on the
  null plane is the pullback  of $\dir$ to this surface. Expressing the
 derivatives  $\partial/\partial x^{\mu}$ in terms of
  $\partial_{i}\equiv \partial/\partial \zeta^{i}$ and substituting them into
 ~(\ref{null12}) we get
\begin{equation}\label{null13c} \dir_{\rm
null}=\Gamma_{i}\partial_{i},
\end{equation}
where
 $\Gamma_{i}=\gamma_{i}+\gamma_{i+2}.$ As a result of $\{
\Gamma_{i},\Gamma_{k} \}=0$,  we find that $\dir_{\rm null}^{2}=0$ and
then the Dirac equation takes the form
$$\dir_{\rm null}\psi=0,$$
which
should be $Sp(1,{\bf R})$ invariant. If so, this Dirac equation is physically
sensible one.  But we can prove more strong proposition that the Dirac
operator is invariant under the groups  $Sp(1,{\bf R})$ and $\cal D$.
. Because of  $\Pi_{i}=\partial_{i}$, we can write
down~(\ref{null13c}) as
\begin{equation}\label{null14}
\dir_{\rm null}=\Gamma_{i}\Pi_{i}
\end{equation}
for the case of {\it the scalar
representation} of the generators.  The generators of the desired $Sp(1,{\bf
R})$ spinor representation are given by ~(\ref{null6a}), where the
generators of the corresponding $SO (2,2)$ representation are substituted.
Then, using the properties of $\gamma$-matrices we find that $\Gamma_i$ are the
tensor operators under $Sp(1,{\bf R})$ and $\cal D$:
\begin{equation}\label{null15}
{[}\Gamma_{i},X_{a}^{(s)}{]}=-(\sigma_{a})_{ki}\Gamma_{k} \quad,\quad
{[}\Gamma_{i},D^{(s)}{]}=\Gamma_{i}.
\end{equation}
Using Eq.~(\ref{null10}) we get
$$[\dir_{\rm null},X_{a}^{(l)}+X_{a}^{(s)}]=
[\dir_{\rm null},D^{(l)}+D^{(s)}]=0,$$
where $X_{a}^{(l)}$ and $D^{(l)}$ are the generators of the scalar
representations of $Sp(1,{\bf R})$ and $\cal D$. The explicit form of
these generators is not essential because the commutation
relations~(\ref{null10}
) contain  all needed information. Thus, the existence of the invariant
Dirac equation~(\ref{null13c}) is proved.

\paragraph{3}
By analogy with the previous case we can consider the null planes in the
Minkowski space with  $\eta_{\mu\nu}=\mbox{diag} (+1,-1,-1,-1)$. But now,
 the vector  $n_1$ is a lightlike vector whereas  $n_2$ is a space-like one:
\begin{equation}\label{null17}
n_{1}\cdot n_{1}=n_{1}\cdot n_{2}=0 \quad,\quad
n_{2}\cdot n_{2} =-1.
\end{equation}
Let us choose them as
$$n_{1}^{\mu}=(1, 1, 0, 0) \quad,\quad
n_{2}^{\mu}=(0, 0, 0, 1).$$
The relations~(\ref{null1}),(\ref{null2}),(\ref{null6}) and (\ref{null12}
) remain invariable. One can conclude that the subgroup of the $SO(3,1)$
group defined by ~(\ref{null5}) is a three-parametric group with the
generators
\begin{equation}\label{null17a}
S=J_{02}+J_{12} \quad,\quad A_{1}=J_{01} \quad,\quad A_{2}=J_{03}+J
_{13}
\end{equation}
which form the algebra
$$[A_{1},A_{2}]=-A_{2} \quad,\quad [A_{1},S]=-S \quad,\quad [A_{2},S]
=0. $$
The generator  $S$ does not change the vectors $n_i$ and may be omitted.
The rest generators form the  group  $\cal H$ which is a solvable one.
The generator $A_1$ produces dilatations of the vector $n_i^\mu$ and the
generator $A_2$
 shifts $n_2^\mu$ by the light-like vector  $n_1^\mu$:
$$(A_i )^\mu_{\ \nu}n_k^\nu =-\delta_{ik}n_1^\mu.$$
The translation generators  $\Pi_{1}=P_{0}+P_{1}$ and $\Pi_{2}=P_{3}$
are the tensor operators under $\cal H$:
\begin{equation}\label{null18}
[\Pi_{i},A_{k}]=\delta_{ik}\Pi_{1}.
\end{equation}
Choosing the null plane coordinates so that
the equality $\Pi_{i}=\partial_{i}$ remains correct and considering the
 pullback of $\dir$ to this null plane we obtain $\dir$ in the
form~(\ref{null14}), where
$\Gamma_{1}=\gamma_{0}-\gamma_{1}$ and $\Gamma_{2}=-\gamma_{3}$. Using
 these  $\Gamma-$matrices we can observe that the Dirac operator
squared  $\dir_{\rm null}^2$ is equal to
$$\dir_{\rm null}^2=- \partial_{2}^2 $$
and it is not invariant under the holonomy group $\cal H$ of the null plane.
Indeed, this follows from the infinitezimal transformation of the holonomy
group (\ref{null18})
$$\delta_{\epsilon}\dir_{\rm null}^2 =
2\Pi_{2}\Pi_{1}\epsilon_{k}\delta_{k2},$$
where the equality  $\partial_{2}=\Pi_{2}$ was taken into account. The
Dirac operator $\dir_{\rm null}$ is also non invariant under $\cal H$,
because
$$[\dir_{\rm null},A_{i}^{(l)}+A_{i}^{(s)}] \not= 0.$$
The non invariant character of $\dir_{\rm null}^2$ and  $\dir_{\rm null}$
is explained by the breakdown of the  $\Gamma-$matrices property to
compose a set of the tensor operators under $\cal H$.

One can consider this fact as an equivalent expression of the impossibility
to introduce the complex structure over a null plane in the  Minkowski
space. Indeed, it is well-known that a spinor field is a representation of
the rotational group. In the above considered case of the null-string in
${\bf R}^{2,2}$ we
 found  the introduction of spinor fields to be possible because the null
 plane symmetry contains the rotation subgroup $SO(2,1)$. This subgroup
exists because each point of the null plane in ${\bf R}^{2,2}$ has  the
two mutually orthogonal and light-like vectors $n_1^\mu$ and  $n_2^\mu$.
The Minkowski signature change, i.e., the passage of ${\bf R}^{2,2}$ into $
{\bf R}^{3,1}$, radically changes the geometric structure of the
null-plane. As a result the introduction of only one light-like basis vector
$n_{1}^{\mu}$~(\ref{null17}) becomes possible, whereas the second linearly
independent vector $n_{2}^{\mu}$ should be a space-like one. These two
vectors can not be transformed one to another by means of a rotation, because
 their norms are different. Therefore, we conclude that the rotation group
 can not be a symmetry of our null-plane in  ${\bf R}^{3,1}$. In place
of the rotation group  $SO(2,1)$ we get a solvable two-parametric group,
as it was proved earlier. Therefore, the signature change makes impossible to
have the rotation group as the null plane symmetry group. This observation
 implies  impossibility to introduce spinor fields and the invariant Dirac
operator over the null plane in the Minkowski space-time.

\paragraph{4}
A change of the null-plane by a light-like worldsheet embedded into
${\Bbb R}^{2,2}$ implies a localization of the $Sp(1,{\Bbb R})$ group
together with the
introduction of the corresponding  worldsheet covariant derivatives. To this
 end let us consider the infinitesimal symplectic transformation with
the   parameters~$\varphi^a$:
\begin{equation}\label{null26}
\Sigma \approx 1+\sigma_{a}\varphi^{a}.
\end{equation}
The characterizing property of the Dirac $\Gamma$-matrices to be invariant
 under the transformations of their vector and spinor indices
\begin{equation} \label{null28}
U\Gamma_{i} U^{-1}=\Gamma_{k}\Sigma_{ki}
\end{equation}
yields the correspondence between the $Sp(1,{\bf R})$ transformations
and the 4-spinor transformations $U$. Using this property  together with
 Eqs.~(\ref{null8a}) and~(\ref{null15}) we  obtain
\begin{eqnarray}\label{null27}
U \approx 1+2X^{(s)}_{a}\varphi^{a}, \\
(\Sigma\sigma_{a}\Sigma^{-1})\, (UX^{(s)a}U^{-1})=\sigma_{a}X^{(s)a}.
\label{null27a}
\end{eqnarray}
The  local $Sp(1,{\bf R})$ transformations ~(\ref{null26}) of the vector $v^i$
 lying on the plane tangent to the string's worldsheet are  ${v'}^i =
\Sigma_k^i v^k$.
Taking into account this transformation law one can define the covariant
differential in the tangent space
 $$Dv^i =dv^i +\omega_k^i (d)v^k , $$
where $\omega_{ik}\in sl(2,{\Bbb R})$ is the connection 1-form in
the tangent space.
 It can be then observed that this covariant derivative transforms under
the local $Sp(1,{\bf R})$ transformations in just the same way as $v^i$.
Then Eqs.~(\ref{null8a}) and~(\ref{null27a}) result in the conclusion that
 the spinor covariant derivative
$${\cal D}_\alpha
\psi=\frac{{\cal D}\psi}{d\zeta^\alpha} \quad,\quad {\cal
D}\psi=d\psi+\omega_k^i (d)(\sigma_a )_{ik}X^{(s)a}\psi $$
correctly transforms under the transformations~(\ref{null26})
\begin{equation}\label{null27b}
{\cal D}'_{\!\alpha }\psi'=U{\cal D}_{\alpha}\psi \quad,\quad
\psi'=U\psi.
\end{equation}
Further we introduce the inverse zveinbein $e_{i}^{\alpha}$ with the
transformation law ${e'}^{\alpha}_{i}=\Sigma_k^i e^{\alpha}_{k}$ under the
local rotations of the worldsheet tangent space~(\ref{null26}).
Equations~(\ref{null27b}) and~(\ref{null28}) yield then the desired
definition of the general covariant worldsheet Dirac operator
$$\dir_{\rm null}=\Gamma_{i}e_{i}^{\alpha}{\cal D}_{\alpha}$$
which correctly transforms under the considered symplectic transformations:
$$\dir_{\rm null}'=U\dir_{\rm null}U^{-1}$$.

\paragraph{5}
Here we study the simplest
model of the bosonic null string embedded in ${\bf R}^{2,2}$.
The null string's  worldsheet has the above desribed symplectic structure.
 In the first place we introduce the local moving frame composed of the
two light-like vectors ($\tn_{i},\tm_{i}$) attached to the null worldsheet:
\begin{equation}\label{null19}
\tn_{i}\cdot\tn_{k}=
\tm_{i}\cdot\tm_{k}=0 \quad,\quad
\tn_{i}\cdot\tm_{k}=2\varepsilon_{ik}
\end{equation}
and next  construct the antisymmetric tensor
$N^{\mu\nu}=\varepsilon^{ik}\tn_{i}^{\mu}\tn_{k}^{\nu}.$  Afterwards,
following to  the prescription
given in~\cite{null-7} we can write down the action
 of the symplectic string in the form
\begin{equation}\label{A}
S=\int d\tau d\sigma\, \Bigl( N_{\mu\nu}\varepsilon^{\alpha\beta}
\partial_{\alpha}x^{\mu}\partial_{\beta}x^{\nu} -E_{ik}\tn_{i}\cdot\tn_{k
}
\Bigr),
\end{equation}
where $\tau=\zeta^{1}\ ,\ \sigma=\zeta^{2}$ and
$E_{ik}$ is a world sheet density ensuring the reparametrization invariance
 of the action $S$. The Euler-Lagrange equations produced by  $S$ are
\begin{equation}\label{B}
\varepsilon^{\alpha\beta}\partial_{\alpha}x^{\mu}
\partial_{\beta}x^{\nu}\varepsilon^{im}
\tn_{m\nu}=E^{ik}\tn^{\mu}_{k}
\end{equation}
\begin{equation}\label{C}
\varepsilon^{\alpha\beta}\partial_{\alpha}N_{\mu\nu}\partial_{\beta}x^{\nu}
=0.
\end{equation}
Let us  look for the general solution of Eqs.~(\ref{B}),(\ref{C}) in the
form of the expansion
\begin{equation}\label{D}
\partial_{\alpha}x=a_{\alpha}^{i}\tn_{i}+b_{\alpha}^{i}\tm_{i}
\end{equation}
  If the expansion~(\ref{D}) is substituted into Eqs ~(\ref{B}) one  yields
 the conditions
\begin{equation}\label{E}
\begin{array}{l}
b^{2}_{\alpha}=\lambda b^{1}_{\alpha} \quad,\quad E_{ik}=\left(
\begin{array}{ll}
1 & \lambda \\
\lambda & \lambda^{2}
\end{array}\right) E_{11} \\
b^{1}_{\alpha}=-\frac{E_{11}}{|a|}(a^{2}_{\alpha}-\lambda a^{1}_{\alpha
})
\quad,\quad |a|=\det a^{i}_{\alpha}\not= 0.
\end{array}
\end{equation}
Taking into account the conditions ~(\ref{E}) we can present the solution
~(\ref{B}) as
\begin{equation}\label{F}
\partial_{\alpha}x=a_{\alpha}^{k}\tn_{k}-\frac{E_{11}}{|a|}(a^{2}_{\alpha}-
\lambda a^{1}_{\alpha})(\tm_{1}+\lambda\tm_{2}).
\end{equation}
It follows from  Eq.~(\ref{F}) that the worldsheet metric $g_{\alpha\beta }$
\begin{equation}\label{G}
g_{\alpha\beta}\equiv (\partial_{\alpha}x \cdot\partial_{\beta}x)=
\frac{E_{11}}{|a|}(a^{2}_{\alpha}-\lambda
a^{1}_{\alpha})(a^{2}_{\beta}-\lambda a^{1}_{\beta}).
\end{equation}
 induced by the embedding into  ${\bf R}^{2,2}$ is  degenerate, i.e.,
$|g_{\alpha\beta}|=0, $ as it should be for the light-like surface.
Moreover, the factorization  of $g_{\alpha \beta}$~(\ref{G}) shows that it
 can be vanished by the choice $a_{\alpha}^{2}=\lambda a_{\alpha}^{1}$.
 In view of these  conditions  the moving frame $(\tn_{i},\tm_{i})$
orientation concerning the null string's  worldsheet and its embedding into
${\bf R}^{2,2}$  are fixed.
The solution of the string equations~(\ref{B}) implied by the conditions
$g_{\alpha\beta}=0$, or equivalently $|a|=0$, takes the form
\begin{equation}\label{null33}
\partial_{\alpha}x=e_{\alpha}^{\tilde{a}}n_{\tilde{a}} \quad,\quad
E_{11}=-2|e_{\alpha}^{\tilde{a}}| \quad,\quad g_{\alpha\beta}=0,
\end{equation}
where the isotropic vectors $n_{\tilde{a}}$
($\tilde{a}=\tilde{1},\tilde{2}$) tangent to the null string's worldsheet
and zveinbein $e_{\alpha}^{\tilde{a}}$
are defined by the relations
\begin{equation}\label{null34}
n_{\tilde{a}}=
\left(
\begin{array}{l}
n_{\tilde{1}} \\ n_{\tilde{2}}
\end{array}
\right)=
\left(
\begin{array}{l}
\tn_{1}+\lambda \tn_{2} \\
\tm_{1}+\lambda \tm_{2} \\
\end{array}
\right) \quad,\quad
e_{\alpha}^{\tilde{a}}=(e_{\alpha}^{\tilde{1}}, e_{\alpha}^{\tilde{2}})
=(a_{\alpha}^{1},b_{\alpha}^{1})
\end{equation}
and the conditions
\begin{equation}\label{null28a}
e_{\tilde{a}}^{\alpha}e_{\alpha}^{\tilde{b}}=
\delta_{\tilde{a}}^{\tilde{b}} \quad,\quad
e_{\tilde{a}}^{\alpha}e_{\beta}^{\tilde{a}}=\delta_{\alpha}^{\beta}
\quad,\quad e_{\tilde{a}}^{\alpha}= (\det e_{\alpha}^{\tilde{a}})^{-1}
\varepsilon_{\tilde{a}\tilde{b}}
\varepsilon^{\alpha\beta}e_{\beta}^{\tilde{b}}.
\end{equation}
Taking into account the solutions~(\ref{null33}),(\ref{null34}) we get
$N_{\mu\nu}\partial_{\beta}x^{\nu}=-2e_{\beta}^{\tilde{2}}n_{\tilde{1}}
$.
Substituting the above expression into Eqs.~(\ref{B})
we reduce them  to the  equations
\begin{equation}\label{null35}
\varepsilon^{\alpha\beta}\partial_{\alpha}
(e_{\beta}^{\tilde{2}}n_{\tilde{1}})=0.
\end{equation}
 Using the solutions~(\ref{null33}) one  can present the 4-vector
$n_{\tilde{1}}$  in the form $n_{\tilde{1}}=e_{\tilde{1}}^{\gamma}
\partial_{\gamma}x.$
 The next substitution of this representation into Eqs.~(\ref{null35})
transforms the latter to
\begin{equation}\label{null36}
\varepsilon^{\alpha\beta}\partial_{\alpha}\left[
e_{\beta}^{\tilde{2}}e_{\tilde{1}}^{\gamma}\partial_{\gamma}x\right]=0.
\end{equation}
In view of the arbitrariness in the choice of the worldsheet gauge
we can fix it by the conditions $e_{\tau}^{\tilde{2}}=0\ ,\
e_{\sigma}^{\tilde{2}}=e_{\tau}^{\tilde{1}}$. In this gauge
 Eqs.~(\ref{null36})take the form
$$\stack{..}{x}^{\mu}=0 \Longrightarrow x^{\mu}=q^{\mu}(\sigma)+{\cal
 P}^{\mu}
(\sigma)\tau,$$
which coincide with the equations of null string embedded in the
Minkowski space~\cite{Shi,Zpb}. By the virtue of
the symmetry between $\tau$ and $\sigma$ for this  embedding
 into ${\bf R}^{2,2}$ there exists another  gauge
 $e_{\sigma}^{\tilde{2}}=0$\ ,\ $e_{\sigma}^{\tilde{1}}=
e_{\tau}^{\tilde{1}}$,
in which the string equations~(\ref{null36}) take the form
$$\stack{\prime\prime}{x}^{\mu}=0.$$
 Thus, we conclude that the null string dynamics described by the action~
(\ref{A}) in ${\bf R}^{2,2}$ is selfconsistent.

The following natural step is the introduction of the worldsheet spinor
fields into the bosonic action~(\ref{A}). This extension is supported by
the existence of the invariant worldsheet Dirac operator, as it was shown
 here. The simplest extension consists in the addition of the kinetic term
 $S_{\rm fermi}$ for the massless Dirac field
  $$S_{\rm fermi}= \lambda\int d^2 \zeta\, E \overline{\psi} \dir_{\rm
null}\psi,$$
where $E$ is the world sheet density contructed of
$E_{ik}$~\cite{Zpb,null-7}. Note, that this extension doesn't suggest the presence
of any supersymmetry.
 The inclusion of spinor fields may shed new light onto the troubles of
the string theory, in particular, the problem of the mass/tension generation
 since the worldsheet spinor structure doesn't depend of the tension value.

Other generalizations such as the  supersymmetrization of the world sheet
 or/and ${\Bbb R}^{2,2}$ space, introduction of the worldsheet
electromagnetic or other internal fields, embedding into curved space and
other backgrounds are also  possible.

We would like to thank  S.Ketov and U.Lindstr\"om for useful discussions.
  A.Z. thanks FYSIKUM, where part of this work was done, for the kind hospitality.
This  work was supported in part Ukrainian SFFI grants F4/1751 and F/1794.
 A.Z. is supported by the grant of the Royal Swedish Academy of Sciences.


\begin{thebibliography}{}

\bibitem{Shi}  A. Shild. Phys. Rev. D 16 (1977) 1722;\\
 A.Karlhede, U.Lindstr\"om. CQG 3 (1986) L73;\\
 F.Lizzi, B.Ray, G.Sparano and G.Srivastava. Phys. Lett. B 182 (1986) 326
=2E
\bibitem{Zpb} A.A.Zheltukhin. JETP. Lett. 46 (1987) 208; Soviet J. of
Nucl.Phys. 48 (1988) 587.
\bibitem{null-7} A.A.Zheltukhin, Teor. Mat. Fiz. 77 (1988) 377.
\bibitem{Ba-Zh} I.A. Bandos and A.A.Zheltukhin. Phys. Lett. B261 (1991) 245;
Fortschr. Phys. 41 (1993) 619.
\bibitem{Il-Zh} A.A. Zheltukhin. Phys. Lett. B233 (1989), 112;\\  K. Ilienko
and A.A. Zheltukhin. Clas. Quant. Grav. 16 (1999) 383.

\bibitem{Ro-Zh} A.A. Zheltukhin.  Clas.Quant.Grav.12 (1996), 2357;\\ S.N.
  Roshchupkin and A.A. Zheltukhin. Nucl. Phys. B 543 (1999) 365;
 hep-th/9806054,

\bibitem{Tw1} P.K. Townsend. Phys. Lett. B 277  (1992) 285.
\bibitem{BLT} E.Bergshoeff, L.A. London and P.K. Townsend. Clas.Quant.Grav.
 9 (1992) 2545.

\bibitem{deV-Sa} H.J. De Vega and N. Sanchez. Phys. Lett.B197 (1987) 320;

 Nucl. Phys. B309 (1988) 577.
\bibitem{Un-Ul} U. Lindstr\"om and R. von Unge. Phys. Lett. B 403 (1997)
233.
\bibitem{Bank-} T. Banks, W. Fischler, S.H. Shenker and L. Susskind.
 Phys.Rev. D 55 (1997) 5112.
\bibitem{Yang-}H.S Yang, I. Kim, B-H. Lee. Phys.Rev. D 58 (1998), 085018.

\bibitem{SGNSW} M.Cederwall, A. von Gussich, B.E.W.Nilsson,P.Sundell and
A. Westerberg, The Dirickhle super p-branes in ten-dimensional type IIA and
 IIB supergravity,  hep-th/9611159.
\bibitem{BT} E.Bergshoeff and P.K. Townsend. Super D-branes revisited,
hep-th/9804011.
\bibitem{Vech-} M.Ademollo, L. Brink,A.D'Adda, R.D. Auria, E Napolitano,
S. Scuito, E. Del Guidice, P. Di Vecchia, S. Ferrara, F. Gliozzi, R. Musto,
 R. Pettorino and J. Schwarz. Nucl. Phys. B 111 (1976) 77.
\bibitem{Fra} A.D'Adda, F.Lizzi. Phys. Lett. B 191 (1987) 85;\\
E.S. Fradkin and A.A. Tseytlin. Phys. Lett. B 106 (1981) 63.

 \bibitem{null-1} H.Ooguri and C.Vafa. Mod.Phys. Lett A 5(1990), 1389;
 Nucl.Phys. B 361 (1991) 469.
\bibitem{Bel-} A.A. Belavin, A.M. Polyakov, A.S. Schwartz and Y. Tyupkin.
 Phys.  Lett. B59 (1975) 85;\\ M.F. Atiyah and R.S. Ward. Commun. Math. P
hys. 55 (1977) 117;\\ E.F. Corrigan, D.B. Fairlie, R.C. Yates and P. Goddard.
 Commun. Math. Phys. 58 (1978) 223;\\ Q-Han Park Phys. Lett. B 238 (1989)
 287.
\bibitem{Ket1} S.V. Ketov, H. Nishino,and S.J. Gates Jr. Phys. Lett. B 307
 (1993) 323; B 307 (1993) 331.

\end{thebibliography}
\end{document}